%% file: icse25.tex
\documentclass[10pt,conference]{IEEEtran}
\IEEEoverridecommandlockouts

\usepackage{comment}
\usepackage{cite}
\usepackage{amsmath,amssymb,amsfonts}
\usepackage[linesnumbered,ruled,vlined]{algorithm2e}
\SetAlgoNlRelativeSize{0}
\SetNlSty{textbf}{}{}
\usepackage{graphicx}
\usepackage{textcomp}
\usepackage{xcolor}
\usepackage{enumitem}
\usepackage{balance}
\usepackage{subfigure}
\usepackage{multirow}
\usepackage{mdframed}
\usepackage[most]{tcolorbox}
\usepackage{tabularx}
\usepackage{booktabs}
\usepackage{soul}
\usepackage[]{collab}
\usepackage{ulem} 
\usepackage{url}
\usepackage{hyperref}
\usepackage{ulem}
\usepackage{algpseudocode}
\usepackage[normalem]{ulem} 

\usepackage[font=normalsize,labelfont=bf,format=plain]{caption}

\usepackage{tikz}

\newcommand*\circled[1]{\tikz[baseline=(char.base)]{
            \node[shape=circle,draw,inner sep=0.5pt] (char) {#1};}}

\def\BibTeX{{\rm B\kern-.05em{\sc i\kern-.025em b}\kern-.08em
    T\kern-.1667em\lower.7ex\hbox{E}\kern-.125emX}}

\newcommand\name{COCA\xspace}

\newcommand{\ie}{{\textit{i.e.}},\xspace}
\newcommand{\eg}{{\textit{e.g.}},\xspace}
\newcommand{\fixedwidth}[1]{{\ttfamily \small #1}}

\definecolor{ballblue}{rgb}{0.13, 0.67, 0.8}
\definecolor{OLGreen}{rgb}{0.239, 0.522, 0.290}
\collabAuthor{jy}{ballblue}{Jinyang Liu}
\collabAuthor{yc}{blue}{Yichen Li}
\collabAuthor{yl}{OLGreen}{Yulun Wu}

\collabAuthor{gb}{orange}{Guangba Yu}
\collabAuthor{zh}{purple}{Zhihan Jiang}
\collabAuthor{jz}{magenta}{Jiazhen Gu}
\collabAuthor{jj}{pink}{Junjie}
\newcommand{\add}[1]{\textcolor{black}{#1}}

\definecolor{mygray}{RGB}{211,211,211} 
\definecolor{morandi}{RGB}{210,225,225} 

\begin{document}
\pagestyle{plain}

\title{COCA: Generative Root Cause Analysis for Distributed Systems with Code Knowledge}



\author{
  \IEEEauthorblockN{
    Yichen Li\IEEEauthorrefmark{2},
    Yulun Wu\IEEEauthorrefmark{2},
    Jinyang Liu\IEEEauthorrefmark{2},
    Zhihan Jiang\IEEEauthorrefmark{2},
    Zhuangbin Chen\IEEEauthorrefmark{3},
    Guangba Yu\IEEEauthorrefmark{2}\IEEEauthorrefmark{4}\thanks{\IEEEauthorrefmark{4}Guangba Yu is the corresponding author.},
    Michael R. Lyu\IEEEauthorrefmark{2}
  }
  \IEEEauthorblockA{\IEEEauthorrefmark{2}The Chinese University of Hong Kong, Hong Kong SAR, China.\\
    Email: \{ycli21, ylwu24, jyliu, zhjiang22, lyu\}@cse.cuhk.edu.hk, yugb5@mail2.sysu.edu.cn}
  \IEEEauthorblockA{\IEEEauthorrefmark{3}School of Software Engineering, Sun Yat-sen University, Guangzhou, China.\\
    Email: chenzhb36@mail.sysu.edu.cn}
}

\maketitle

\begin{abstract}

Runtime failures are commonplace in modern distributed systems. When such issues arise, users often turn to platforms such as Github or JIRA to report them and request assistance. Automatically identifying the root cause of these failures is critical for ensuring high reliability and availability. However, prevailing automatic root cause analysis (RCA) approaches rely significantly on comprehensive runtime monitoring data, which is often not fully available in issue platforms. Recent methods leverage large language models (LLMs) to analyze issue reports, but their effectiveness is limited by incomplete or ambiguous user-provided information.

To obtain more accurate and comprehensive RCA results, the core idea of this work is to extract additional diagnostic clues from code to supplement data-limited issue reports. Specifically, we propose COCA, a \underline{co}de knowledge enhanced root \underline{c}ause \underline{a}nalysis approach for issue reports. Based on the data within issue reports, COCA intelligently extracts relevant code snippets and reconstructs execution paths, providing a comprehensive execution context for further RCA. Subsequently,  COCA constructs a prompt combining historical issue reports along with profiled code knowledge, enabling the LLMs to generate detailed root cause summaries and localize responsible components. Our evaluation on datasets from five real-world distributed systems demonstrates that COCA significantly outperforms existing methods, achieving a 28.3\% improvement in root cause localization and a 22.0\% improvement in root cause summarization. Furthermore, COCA's performance consistency across various LLMs underscores its robust generalizability.

\end{abstract}

\input{sections/01_introduction.tex}

\input{sections/02_background.tex}
\input{sections/03_method.tex}

\input{sections/04_evaluation.tex}

\input{sections/05_results.tex}
\input{sections/06_discussion.tex}
\input{sections/07_related_work.tex}

\input{sections/08_conclusion.tex}

\balance
\normalem
\bibliographystyle{IEEEtran}
\bibliography{icse25}

\end{document}

%% file: sections/01_introduction.tex
\section{Introduction}
\noindent The reliability of complex distributed systems is critical for ensuring seamless operation and user satisfaction. However, despite best efforts, users may still encounter various issues and report them through issue tracking systems (such as JIRA~\cite{JIRA}). These reports often contain essential information such as issue descriptions, runtime logs, and stack traces. It is crucial for developers to identify the root causes and mitigate them promptly to maintain system reliability. However, manually understanding these reports and identifying the underlying root causes can be labor-intensive and error-prone, given the complexity of modern distributed systems. 

Significant research has been devoted to developing automatic root cause analysis (RCA) approaches~\cite{yu2023nezha,li2020swisslog,lin2016log, rosenberg2020spectrum, lin2020fast, nlprca}.
Extensive approaches focus on using comprehensive run-time monitoring data (\eg logs, metrics, and traces) to locate root causes~\cite{lin2016log,huang2024demystifying}.
\add{They typically compare the fault-suffering data collected from the faulty states with the fault-free data collected from the normal state of the system to identify root causes~\cite{yu2023nezha,rosenberg2020spectrum,lin2016log}, }
Nevertheless, this monitoring data is often unavailable in practice because it requires extensive infrastructure to collect the required data. \add{For instance, commonly used issue tracking systems, such as JIRA~\cite{JIRA}, generally contain only issue descriptions and fault-suffering data in reports submitted by users~\cite{ist}.}
\add{Another related work is fault localization, which pinpoints specific locations in the code responsible for software failures. However, bug reproduction-based fault localization techniques~\cite{kang2024quantitative,kang2023explainable,just2014defects4j} face challenges for localizing the faulty code in distributed systems, which is extremely challenging with only issue reports available~\cite{ist, yuan2014simple}. Furthermore, fault localization techniques are limited to addressing code-related issues and cannot diagnose the non-code-related issues, such as network failures.}


To address the limitations of monitor data-driven RCA approaches, \add{recent approaches leverage the semantic comprehension and reasoning capabilities of Natural Language Processing (LLMs) techniques to identify root causes from issue reports, particularly when only limited run-time monitoring data is available ~\cite{RCACopilot,an2024nissist,roy2024exploring,ahmed2023recommending, nlprca}.}
\add{
Unfortunately, the information contained in user-submitted issue reports is often incomplete or ambiguous~\cite{ist, RCACopilot}. Users typically can only provide symptom-related details, such as brief issue descriptions and snippets of log messages, as the complexity of the distributed systems makes it difficult for them to detail the underlying execution logic directly responsible for the issue. This lack of detailed execution information, including the execution logic proceeding to failure, architectural and interaction patterns embedded in system code limits the ability of LLMs to recommend the accurate root causes.}
\add{To bridge this gap, the core idea of this work is to accurately retrieve diagnostic clues (\eg execution logic) from the complicated source code of a distributed system to supplement RCA in data-limited issue reports.
Given the limited information in an issue report (\eg issue description, logs, and stack traces), our approach deals with the problem by reconstructing the execution logic preceding the issue by identifying and analyzing the relevant code.} This allows us to infer root causes even with incomplete textual data in the issue report. Inspired by the capabilities of LLMs to comprehend both code and natural language, our intuition is to leverage LLMs to reason the execution paths prior to issue and understand the logic from relevant code. 

However, it is non-trivial to integrate knowledge from codebase for diagnosing root causes in distributed systems without execution. We have identified three major challenges.
(1) Linking logs directly to specific code positions is difficult because detailed metadata, like line numbers, is often missing, and runtime log messages are usually different from the original logging statements~\cite{jiang2023large,bushong2020matching}.
(2) The complexity of distributed systems, with various invocation mechanisms such as static calls and Remote Procedure Calls (RPCs)\footnote{a widely-used protocol in distributed systems that allows a program to run a procedure on another machine as if it were local.}, makes it hard to piece together execution paths from code positions.
(3) Due to the context window limitations, LLMs struggle to understand long codebases~\cite{peng2023generative,ding2024crosscodeeval}, making it difficult to use LLMs to analyze complex and extensive codebases of distributed systems during RCA.

\textbf{Our work.} 
To address these challenges, we propose \name, the first \underline{co}de knowledge enhanced root \underline{c}ause \underline{a}nalysis approach.
To address the first challenge, the \textit{logging source retrieval} phase employs a static analysis-based backtracking approach to accurately identify the source logging statements, even when the actual log messages differ significantly from the original logging statements.
These logging statements then provide code snippets directly related to the issue reports, such as the lines of code executed immediately before the issue occurs.
Next, we design an \textit{execution path reconstruction} step to incorporate method invocations, thereby providing a more comprehensive execution context prior to the occurrence of an issue. 
Beyond analyzing call graphs between methods, we propose an RPC bridging method, aiming to reconstruct interactions prior to failure via RPCs.
Then, the \textit{failure-related code profiling} phase employs code snippet indexing and retrieval methods to analyze all obtained code snippets, which aims to profile the snippets into a more compact form, 
This allows them to fit within a reasonably sized window suitable for processing by LLMs.
Finally, during the \textit{root cause inference} phase, we construct a prompt based on historical issue reports, the target issue report, and all profiled code knowledge. This prompt is then used to query the LLM, which generates a thorough root cause summary and localized responsible components. 

To evaluate COCA, we conduct a comprehensive evaluation based on the dataset collected from five real-world distributed systems. Our results demonstrate that \name achieves the best performance overall metrics in both root cause summarization and localization. More precisely, \name surpasses the current leading method by 28.3\% in root cause localization (as measured by \textit{Exact Match}) and by 22.0\% in root cause summarization (\eg \textit{BLEU-4~\cite{papineni2002bleu}}). Additionally, \name consistently enhances performance across various underlying LLMs (\eg LLaMa-3.1~\cite{llama31}), thereby affirming its robust generalizability. Moreover, we explore the individual contributions of each phase and provide the corresponding reasoning.

This paper’s contributions are summarized as follows:
\begin{itemize}[leftmargin=*, topsep=0pt]
    \item To our best knowledge, \name\footnote{https://github.com/YichenLi00/COCA} is the first work incorporating code knowledge into the automatic root cause analysis framework for issue reports in distributed systems. 
    \item We proposed and implemented \name \add{with multiple generalized novel tools (\eg RPCBridge) which solve several critical challenges in RCA field.}
    \item We collected and labeled a real-world issue dataset of distributed systems for RCA, which includes 106 real-world issues covering five widely used distributed systems.
    \item We extensively evaluate the performance of \name on real-world datasets. The results demonstrate the effectiveness of \name and the adaptability of \name with different backbone models.
\end{itemize}

%% file: sections/02_background.tex
\section{Background and Motivation}\label{sec: background}

\subsection{Background}\label{sec: iterm}

\begin{figure}[t]
    \centering
    \includegraphics[width=0.9\columnwidth]{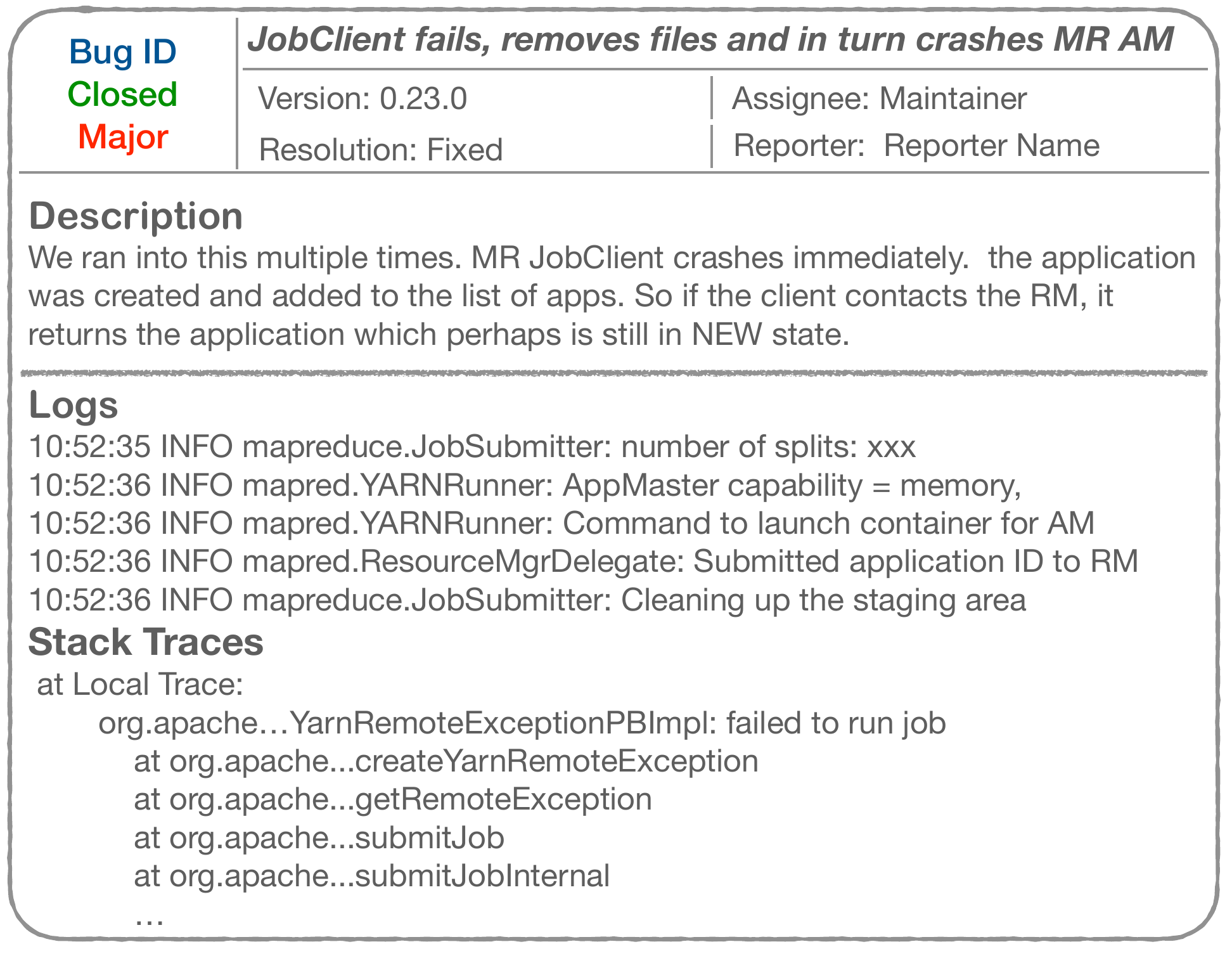}
              \vspace{-0.05in}

    \caption{Example of an issue report.}
              \vspace{-0.15in}
    \label{fig: bugreport}
\end{figure}

\noindent\textbf{Issue Reports in Distributed Systems.} 
Issue reports, including bug reports, GitHub issues, and cloud incident tickets~\cite{chen2020towards, liu2023incident}, are reported by users to track and manage various types of system problems. We unify the terms throughout the paper to issue reports for distributed systems.

As shown in Figure~\ref{fig: bugreport}, when users encounter a system issue (\eg JobClient fails), they typically report the problem on the system's issue tracking platform as an issue report. An issue report usually contains attributes such as the \textit{issue title}, \textit{description}, \textit{logs}, \textit{stack trace}, and \textit{meta information} (including version, severity, and status). The \textit{description} is a free text written by users that describes the problem, aiding in understanding the issue. \textit{Logs} offer a detailed record of system events and actions preceding the failure. Each log message typically includes a timestamp indicating when the event occurred, a text template, and dynamic variables that provide context-specific information (\eg request IDs). Log sequences can reflect the execution path of the system within a specific time frame~\cite{huo2023autolog, chen2018automated,liu2023scalable}. For instance, the provided logs in the issue report suggest that the JobClient failure occurred after posting the application to RM and the staging area clean up. \textit{Stack traces} act as a snapshot of the call stack at the point where a failure occurred to pinpoint the context of the failure.

Addressing an issue report manually can be a time-consuming and tedious procedure~\cite{yuan2010sherlog, yuan2014simple,ist,huang2024faultprofit}. It requires rich domain knowledge from maintainers to understand, reproduce, and eventually fix the problem. This difficulty is further amplified in complex distributed systems.

 \begin{figure*}[tbp]
  \centering
     \includegraphics[width=1.9\columnwidth]{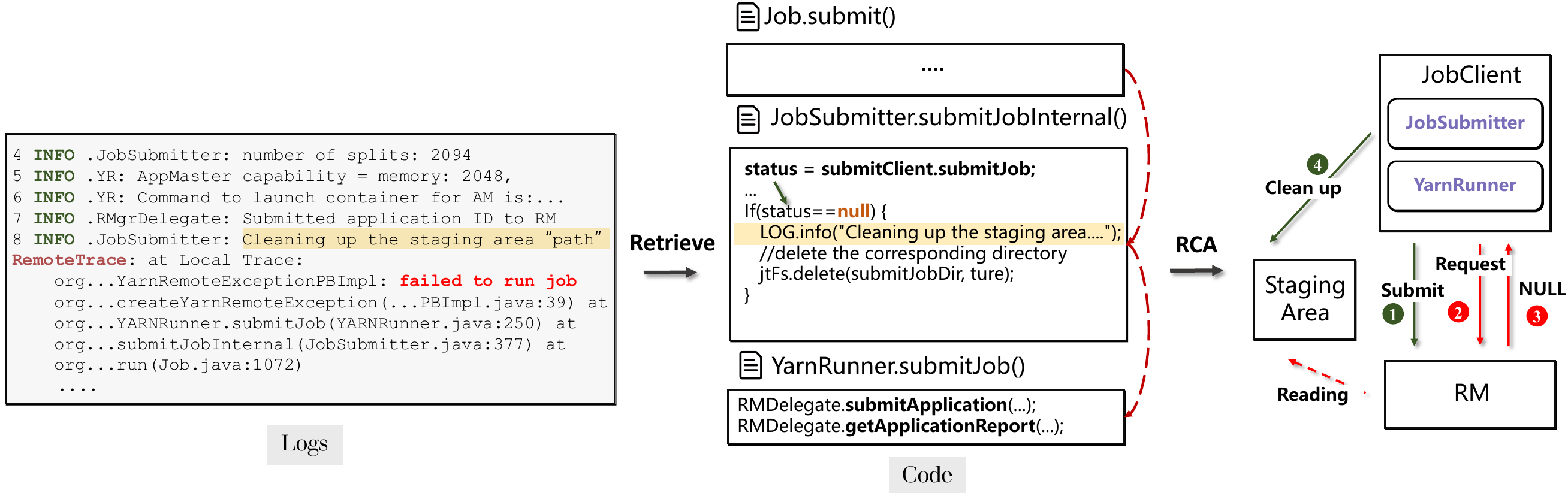}
     \caption{A motivating example: MapReduce-2953.}
     \label{fig:case}
     \vspace{-0.2in}
 \end{figure*}


\subsection{Motivating Study}\label{sec: case}


In this section, we present a motivating example to demonstrate our key insight: using code knowledge to enhance system failure diagnosis. 
Specifically, we present a real-world issue report, MAPREDUCE-2953\footnote{\url{https://issues.apache.org/jira/browse/MAPREDUCE-2953}}, from the widely-used distributed system MapReduce~\cite{dean2008mapreduce}. 

A MapReduce job fails immediately after the user submits the job from the client machine.
To diagnose this issue, a common practice is to investigate the error log messages to understand what went wrong. For example, the stack trace indicates that this job failed after submission. 
Once this error is identified, the next step is to trace back through the logs to find preceding events that might lead to the issue. In this case, line 8 is particularly suspicious because a \texttt{cleaning up} action is performed after the job is submitted.

While log analysis sheds some light, it falls short of revealing the underlying cause of the failure.
To find out the rationale behind this problem, we need to understand the code execution logic that occurs before the issue arises.
Specifically, \circled{1} the client submits an application to the Resource Manager (RM), and \circled{2} immediately requests an application report. However, at the time of the request, \circled{3} the RM has not completed setting up the application in the staging area (\ie a temporary workspace to organize the necessary resources to run an application), returns the NULL to the client. Due to the incomplete setup, \circled{4} the client then mistakenly perceives the application as removed and initiates a cleanup process, while printing the log message ``\texttt{Cleaning up the staging area...}''. This action occurs while the RM is still reading the staging area after receiving the application submitted from the client. This race condition results in a failed job run.

In this case, existing solutions~\cite{RCACopilot,roy2024exploring,li2020swisslog}, which solely rely on the analysis of reports, exhibit a significant limitation: they can only confirm job failure and pinpoint the suspicious cleanup action. However, accurately determining the root causes without a comprehensive understanding of the job's execution is hard, without the involvement of the relevant code.

To address this limitation, we introduce an innovative approach for a more thorough RCA. Our approach synergies runtime log data with the corresponding executed code, thereby providing a deeper comprehension of job failures. 
Specifically, we plan to leverage the capabilities of LLMs, which have demonstrated proficiency in comprehending both incident tickets~\cite{roy2024exploring,RCACopilot} and code~\cite{yang2023code,gao2023what}, to enhance the issue diagnosis process.

\subsection{Challenges}
While the integration of code knowledge into the RCA procedure seems intuitive, it poses three main challenges:

\textbf{Challenge 1: Sourcing Log Messages.} 
While stack traces provide specific \textit{class names}, \textit{file names}, and \textit{line numbers} for precise location, sourcing the original code lines of log messages is complex~\cite{bushong2020matching}. Ideally, we could use the above mentioned metadata (\eg~ \textit{line numbers}) provided by logging libraries for precise localization. However, obtaining such detailed metadata is usually not feasible in real-world scenarios~\cite{he2020loghub,jiang2024large, wang2022spine}. Furthermore, runtime log messages often differ significantly from the logging statements in the code~\cite{bushong2020matching,schipper2019tracing}, making it difficult to for a precise match. In particular, the variable in the logging statement may consist of multiple dynamic variables across different branches~\cite{huo2023autolog, yuan2010sherlog}.

\textbf{Challenge 2: Reconstructing Execution Paths.} The complexity of execution paths in distributed systems makes their reconstruction from logs and stack traces in issue reports a formidable task~\cite{li2022intelligent, chen2018automated}. Building the dependencies between these code pieces is highly complex due to the diverse nature of dependencies in distributed systems, including static calls, gRPC calls, and others. These complexities block the accurate determination of the code executed prior to failures.

\textbf{Challenge 3: Profiling Failure-Related Code Snippets.}
Assuming the ability to reconstruct the execution paths prior to failures, this would involve a vast amount of code. Failure-related code knowledge comes not only from the executed code but also from its dependencies~\cite{li2024enhancing}. The absence of these dependencies could render code slices hard to comprehend. However, including all executed and dependent code significantly expands the context length, potentially exceeding the model’s input limit, leading to high costs and model confusion.



%% file: sections/03_method.tex
\section{Methodology}\label{sec: method}
 \begin{figure*}[tbp]
  \centering
     \includegraphics[width=2\columnwidth]{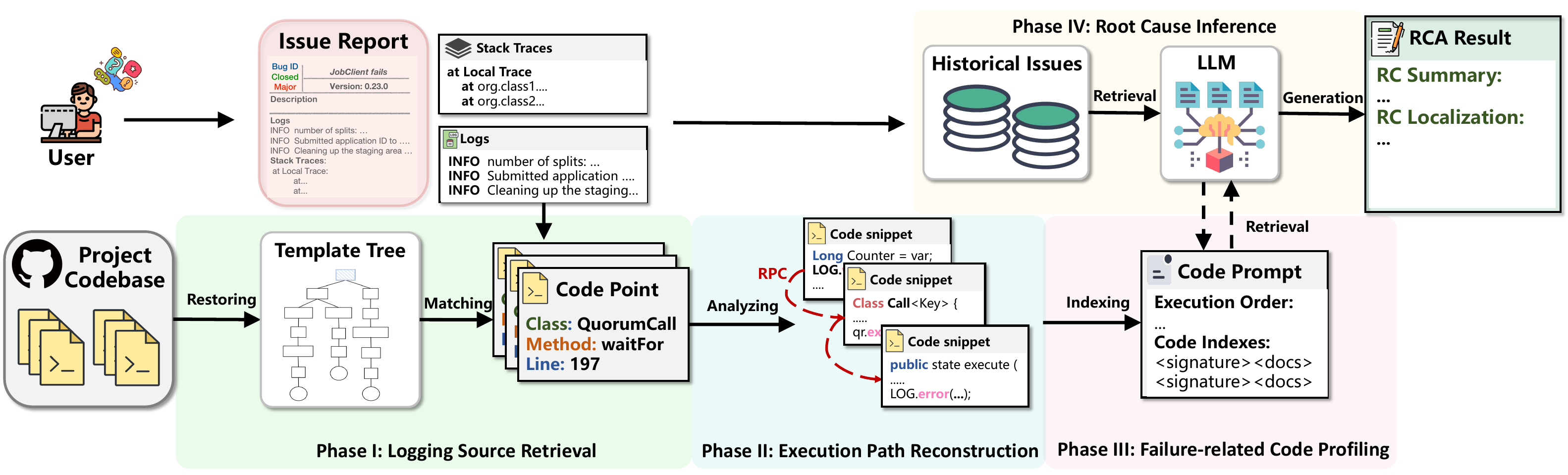}
          \vspace{-0.05in}
     \caption{The workflow of \name.}
     \label{fig:framework}
     \vspace{-0.1in}
 \end{figure*}

\subsection{Overview}\label{sec: method_overview}
To address the above challenges, we propose \name, a framework that enhances RCA of issue reports in distributed systems through code knowledge.

As illustrated in Figure~\ref{fig:framework}, \name takes the issue report along with the associated project code of the corresponding version as input and generates the diagnostic result through a four-phase process. (\uppercase\expandafter{\romannumeral1}) In \textbf{Logging Source Retrieval} phase, \name localizes the code positions (\ie the line numbers of the corresponding logging statements) of log messages in issue reports and maps out the code points spreading across the systems. (\uppercase\expandafter{\romannumeral2}) In \textbf{Execution Path Reconstruction} phase, the identified code points are used to reconstruct the execution paths leading up to the failure with a call graph patched with RPC edges and inter-method analysis. (\uppercase\expandafter{\romannumeral3}) In \textbf{Failure-related Code Profiling} phase, \name analyzes and indexes the extracted method-level code snippets with execution orders with signatures and documents. The LLM can retrieve the full method code of the code snippet based on indexes and initial comprehension of the issue. (\uppercase\expandafter{\romannumeral4}) During \textbf{Root Cause Inference} phase, in conjunction with similar historical issue reports, \name diagnoses the issue using all acquired knowledge and outputs a comprehensive root cause summary and components.

\subsection{Logging Source Retrieval}

\begin{figure}[t]
    \centering
    \includegraphics[width=0.85\columnwidth]{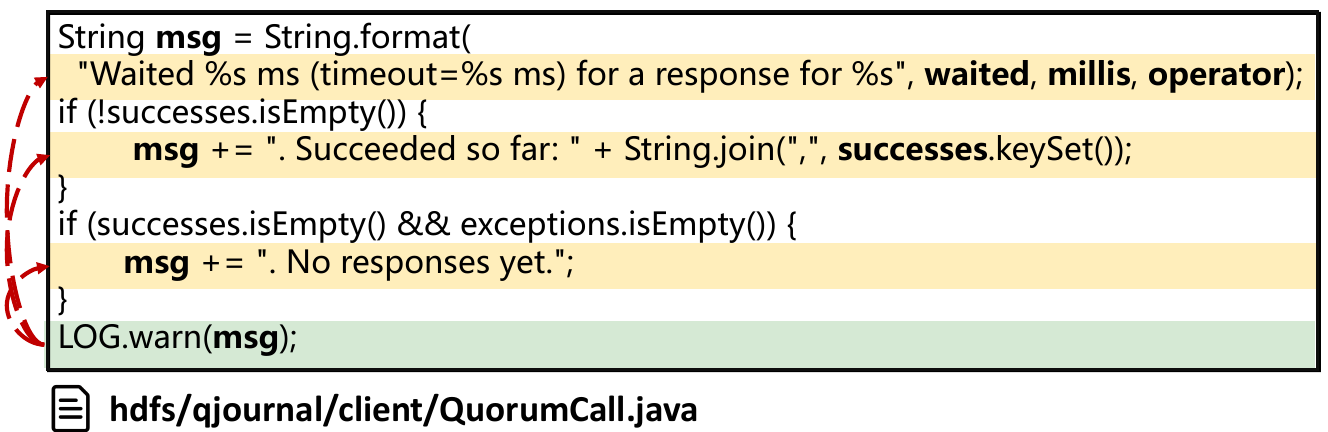}
    \vspace{-2pt}
    \caption{An example of logging statement restoring.}
    \vspace{-10pt}
    \label{fig:restore}
\end{figure}

To address the \textbf{Challenge 1} and identify the precise logging statement sources from the code of log messages in the given issue report, \name employs a two-step approach after execrating all logging statements and their corresponding code positions, which is achieved by analyzing the project's logging library~\cite{li2023exploring}. The first step involves restoring the logging statements that contain constructed variables~\cite{huo2023autolog} back to primitive. For the second step, we introduce a template match algorithm to match log messages with restored logging statements of the corresponding system version.

\subsubsection{Restoring Logging Statements} This step involves restoring logging statements by resolving constructed variables (\eg~\fixedwidth{Log.warn(msg);}), which is further used for matching log messages to their corresponding logging statements. As shown in Figure~\ref{fig:restore}, the raw logging statement only indicates the constructed variable \textit{msg}, instead of the primitive template with constant strings and original variables.

To tackle this issue, we analyze data dependency and control flow to obtain the primitive template set. Inspired by previous work \cite{schipper2019tracing}, \name employs a backtracking approach based on static analysis to identify and restore variables' paths, allowing \name to restore logging statements and obtain uncompressed string constants and variables from a simple \textit{msg}. Specifically, variable \textit{msg} is constructed through several steps via conditional branches. \name restores \textit{msg} back to its original constant strings and variables for further matching. Specifically, after restoring, the logging statement in Figure~\ref{fig:restore} is reinstated to four different primitive templates (\eg~\fixedwidth{Waited \texttt{<*>} ms (timeout=\texttt{<*>}) for a response for \texttt{<*>}. No responses yet.}) for each control branch.

\subsubsection{Log Template Matching}

\input{algorithms/template_match}

Upon restoring all the primitive log templates, the subsequent step involves matching log messages to templates for determining the code positions.

During the process of log message matching, a single log message could potentially be matched to multiple logging templates~\cite{jiang2024large,jiang2023lilac}. For instance, there are two templates $T_1$: (\fixedwidth{"Job created, executing "} + \texttt{<*>}) and $T_2$: (\fixedwidth{"Job created, executing "} + \texttt{<*>} + \fixedwidth{" on node "} + \texttt{<*>}). Every log message generated from template $T_2$ could be potentially matched by $T_1$. This might be due to the fact that for some complexly constructed variables (\eg \textit{msg}), it is challenging to restore them thoroughly, leaving some unsolved variables to have such a general template.

To address this issue, we proposed a template matching Algorithm~\ref{alg:1} to match log messages back to logging statements accurately. Different from general logging parsing problems~\cite{huang2024ulog,he2017drain}, \name performs a non-pruneable recursive matching to review all templates within the prefix tree, maintaining a global perspective. When a log message can match several log templates, \name prioritizes the template with the most static parts (\ie non-\texttt{<*>} characters). \add{Noted that \name currently only analyzes logs from the system being diagnosed without analyzing logs from third-party libraries, as \name focuses on helping maintainers of target systems.}

\subsection{Execution Path Reconstruction}
After successfully matching the code positions of log messages in a given issue report, the subsequent task is to figure out how to incorporate these code points into the RCA process. An intuitive approach is extracting the surrounding lines of logging statements to profile the logs. However, this method presents challenges when applied to distributed systems, where logs are not densely populated~\cite{yuan2014simple,ist,zhao2014lprof}. For two adjacent log messages, numerous methods might execute across different components. Merely extracting surrounding lines without reconstructing and incorporating the executed code may miss important system runtime information~\cite{ist,jia2017logsed,huo2023autolog,li2024go}.

In order to comprehend the pre-failure execution state of the system, we need to address the \textbf{Challenge 2}: reconstructing the execution paths prior to failure based on identified code points. Following the previous works~\cite{huo2023autolog, li2024go, wu2023understanding, chen2018automated}, \name constructs the \textit{inter-procedure control flow graph} (ICFG) for connecting the code points with execution code paths. Typically, the construction of the ICFG involves both \textit{inter-method} and \textit{intra-method} analysis. \textit{Inter-method} analysis primarily focuses on identifying the call graph of the system, which includes determining the relationships of invocations between different methods, both direct and indirect. Conversely, \textit{intra-method} analysis resolves the control flow within each individual method to identify its potential execution paths. 




Additionally, in distributed systems, simply analyzing the call graph to model the invocation relationships is insufficient, given the widespread use of RPC~\cite{Zhang2022CRISP}. RPC is a prevalent mechanism for facilitating interactions between individual instances~\cite{fang2023rpcover}. These calls are made dynamically (\ie via network utilities), making it hard to statically capture the call edges between instances. Worse still, existing approaches leave this issue unsolved~\cite{huo2023autolog, chen2018automated}. To address this, we propose an RPC bridging method suitable for common RPC frameworks such as gRPC\cite{gRPC} and Thrift\cite{thrift}, to improve the precision of constructing call graphs in distributed systems.

\begin{figure}[t]
    \centering
    \includegraphics[width=0.8\columnwidth]{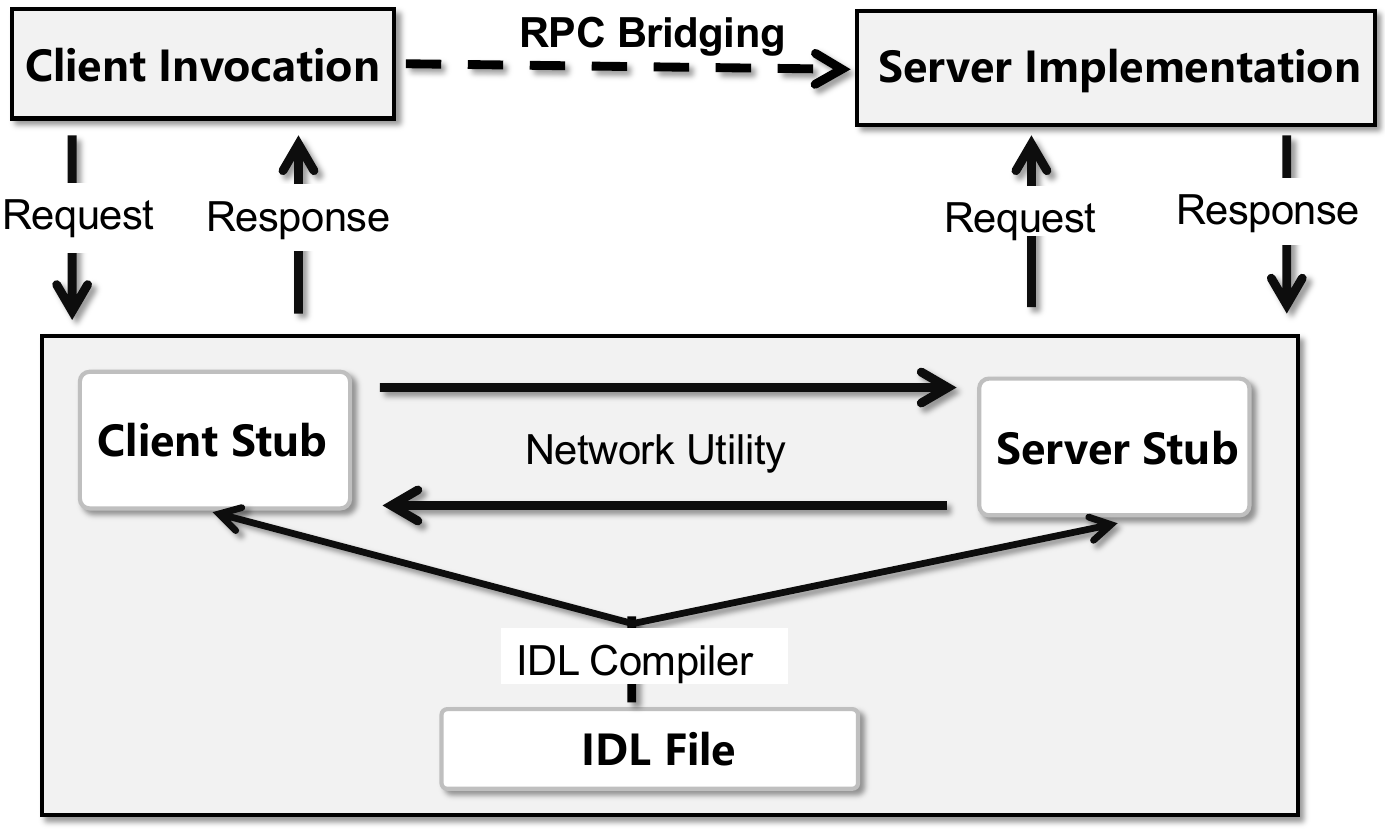}
    \vspace{-2pt}
    \caption{A common RPC call invocation process.}
    \vspace{-15pt}
    \label{fig:gRPC_call}
\end{figure}

Figure \ref{fig:gRPC_call} shows the typical invocation process for a remote call under common RPC frameworks. Initially, RPC functions defined by the Interface Description Language (IDL) are transformed into \textit{client} and \textit{server} stubs. These stubs provide an interface for handling remote calls, including serialization and message sending. Subsequently, the \textit{client} invokes the \textit{client stub}, which ultimately connects to the server stub to call the server’s service functional implementation. In this process, the client stub invokes the server implementation via network utilities. However, as the actual call targets are determined at runtime, this dynamic binding introduces significant uncertainty for static analysis\cite{Li2019TOSEM, Grech2017PTaint, Bruce2020JShrink}, thus breaking the call stack between client and server. 

To address this uncertainty, we leverage the implementation patterns in the call process to bridge the call stack: \name first identifies client invocation to RPC functions by parsing the IDL files (\eg proto files for gRPC and thrift files for Thrift)  to obtain the RPC function names and interface names. \name then searches for matched callees in the call graph, confirming them as client-to-server RPCs. Second, \name extracts the server-side implementation classes, which typically have names similar to their interface counterparts (\eg the interface \textit{ResourceTrack} and its implementation \textit{ResourceTrackService}). We use substring matching for fuzzy identification. To prevent false positives, we further examine the declarations of the matched classes to ensure the implementation of corresponding interface classes. Finally, we replace the callee of client-to-server RPCs with the corresponding server-side implementation functions, completing the client-to-server call stack bridging, as shown by the dashed line in Figure \ref{fig:gRPC_call}.

With this patched call graph and inter-method analysis, \name identifies multiple execution paths based on specified code points for the subsequent step. \add{Note that while \name currently supports gRPC~\cite{gRPC} framework, it can be readily adapted to other RPC (\eg Thrift~\cite{thrift}) frameworks by replacing the code that parses the RPC definition files (i.e., IDL files)}

\subsection{Failure-Related Code Profiling}
Upon reconstructing the execution paths based on code points the failure, \textbf{Challenge 3} emerges: ~\textit{How to deal with thousands of lines of executed code?}

The code profiling phase aims at profiling the code information while concurrently reducing redundant failure-unrelated code snippets and minimizing the cost (\ie context length) of \name. To retrieve the code snippets that contribute to failure comprehension or are prone to failure, and forward them to the backbone LLM, \name utilizes a two-staged approach that involves indexing and retrieval.

\begin{figure}[t]
    \centering
    \includegraphics[width=0.85\columnwidth]{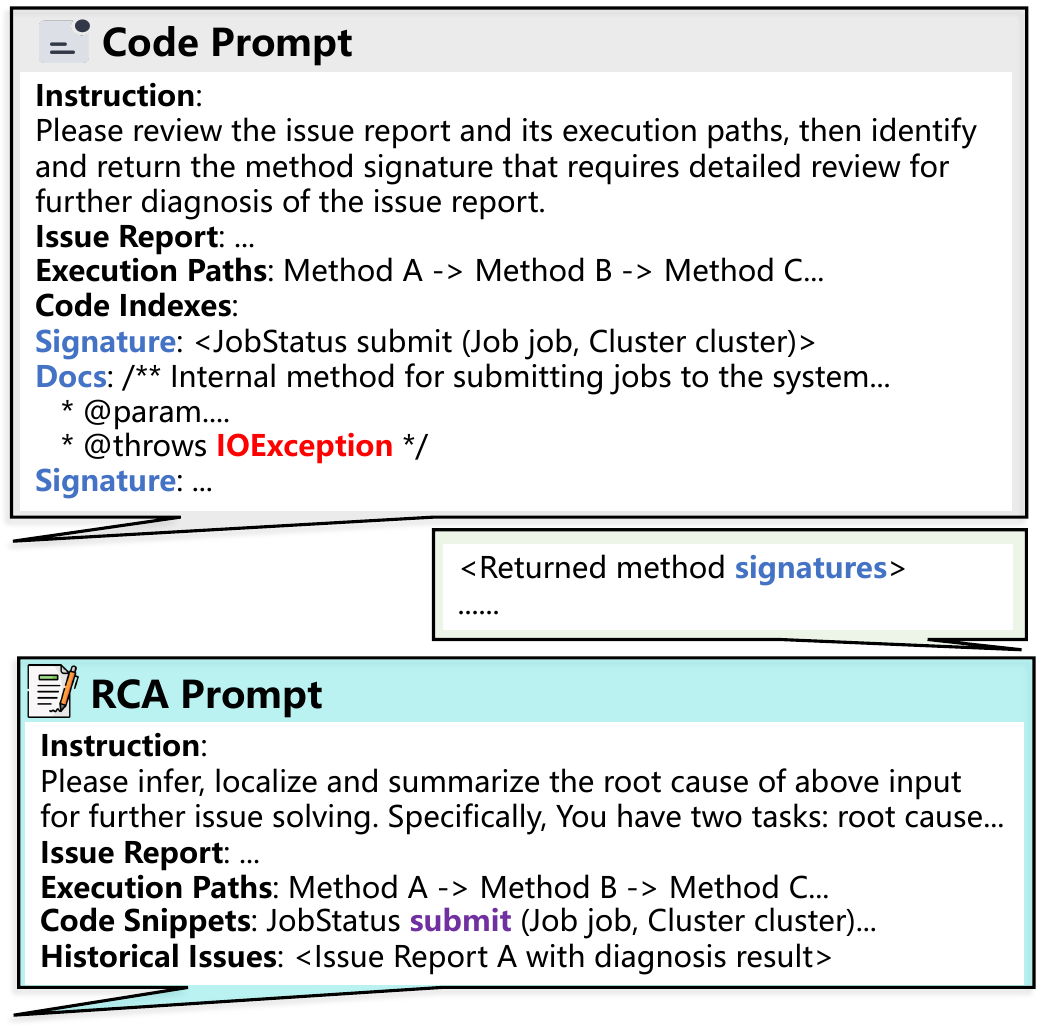}
    \vspace{-0.08in}
    \caption{Example prompt in \name.}
    \label{fig: prompt}
    \vspace{-0.3in}
\end{figure}

\subsubsection{Code Snippet Indexing}
To streamline the process of retrieving relevant code snippets, we implement a method-level indexing framework for executed paths. The indexing framework creates individual indexes for each method in the execution paths, equipping the backbone model with sufficient key information to make selections without the need to process the entire method code. Given the experiment datasets (seen Section~\ref{sec: dataset}) were collected from well-maintained systems, we use standard method docs along with the method signature generated by Soot~\cite{vallee2010soot} as indexes, as shown in code prompt part of Figure~\ref{fig: prompt}. If docs are unavailable, we can generate standard docs using LLM-based methods\cite{Nam2024ICSE, Wang2023EMNLP,li2024enhancing}.

\subsubsection{Code Snippet Retrieval}
We provide the LLM with a comprehensive understanding of the issue report and analyzed execution paths, along with indexes for subsequent retrieval. With the initial comprehension of the target issue report and execution paths preceding to the failure, \name will allow LLM to identify code snippets that contribute to failure comprehension or are prone to failure. To accomplish this, we input both the issue report and the code snippet indexes of all analyzed execution paths into the backbone LLM for retrieval from the codebase. As shown in Figure~\ref{fig: prompt}, it will return the method signatures that require detailed review. Based on these generated method signatures, the corresponding method code is then extracted, preparing for the next phase.

\subsection{Root Cause Analysis}

\subsubsection{In-context Learning}
To learn the historical experiences from previous issue reports of current system, \name employs the in-context learning (ICL) strategy~\cite{gao2023what}, which has proved its effectiveness in bug diagnosis task ~\cite{RCACopilot,roy2024exploring, wang2023rcagent}. The ICL strategy improves the effectiveness of LLMs based on historical experience. Specifically, this strategy provides a few historical issue reports sampled from the collected and labeled dataset, which was detailed in Section~\ref{sec: dataset}

However, the inherent heterogeneity of issue reports presents the challenge~\cite{ist} of finding similar and useful historical examples. Embedding-based similarity metrics, though proved effective for certain applications~\cite{roy2024exploring,RCACopilot}, may not perform optimally in the context of diverse issue reports~\cite{gao2023what}.

We use the BM25~\cite{robertson2004simple} similarity to select these examples. The BM25 function, based on the Term Frequency-Inverse Document Frequency (TF-IDF) method, places greater emphasis on keywords, making it particularly suitable for component matching among issue reports. We specifically select the top five historical issue reports with the highest similarity scores.

\subsubsection{Diagnosis Result Inference}
As illustrated in Figure~\ref{fig: prompt}, after comprehensively analyzing the issue report, reconstructed execution paths, and retrieved code snippets, \name is well-equipped to infer the root cause of the issue.

\name structures the prompt by beginning with the issue report, followed by the relevant knowledge obtained from the analysis that contributes to the root causes analysis. Specifically, \name focuses on two key tasks: root cause summarization and localization. The goal of root cause summarization is to provide a clear and concise guide for maintainers in their subsequent mitigation and fixing efforts. In contrast, root cause localization aims to provide the analyzed scope for the actual root cause, facilitating further fixes by enumerating the potential root cause components ranked by likelihood.

%% file: algorithms/template_match.tex
            

\normalem
\begin{algorithm}[t]
\SetKw{KwRet}{return}
\SetKw{KwRequire}{Input:}
\SetKw{KwEnsure}{Output:}
\SetKw{KwFunction}{Function:}
    \caption{Template Matching Procedure}
    \label{alg:1}
    \KwRequire{\rm Template prefix tree; Log message L;} \\
    \KwEnsure{\rm Matched template $t_{m}$;} 

    \SetKwFunction{FMain}{match}
    \SetKwProg{Fn}{Function}{:}{}
    
    \Fn{\FMain{\rm root, L, results}}{
        \If{\rm len(L) = 0 and root is a leaf node}{
            add template of root to results; \KwRet\;
        }
        \If{\rm L[0] in root.children}{
            \FMain{root[L[0]], L[1:], results}\;
        }
        \If{\rm \texttt{<*>} in root.children}{
            \For{\rm i $\leftarrow$ 0 to len(L)-1}{
                \If{\rm L[i] in root[\texttt{<*>}].children}{
                    \FMain{\rm root[\texttt{<*>}], L[i:], results}\;
                }
            }
        }
    }
    root $\leftarrow$ root node of the template tree\;
    L, results $\leftarrow$ split\_tokens(log message), [\ ]\;
    \FMain{\rm  root, L, results}\;
    $t_{m} \leftarrow$ most\_static\_template(results)\;
    \KwRet{$t_{m}$}\;

\end{algorithm}
\ULforem

%% file: sections/04_evaluation.tex
\section{EXPERIMENT SETUP}\label{sec: evaluation}
\noindent We want to answer the following research questions (RQs).
\begin{itemize}[leftmargin=*, topsep=0pt]
    \item RQ1: How effective is \name in RCA compared with existing methods?
    \item RQ2: What are the impacts of different phases in \name?
    \item RQ3: How generalizable is \name with different backbone models?
\end{itemize}

\subsection{Dataset}\label{sec: dataset}
\begin{table*}[htbp]
    \footnotesize
    \centering
    \caption{Details of the datasets.}
    \label{tab:subjects}
    \begin{tabular}{l>{\raggedright\arraybackslash}p{1.2cm} >{\raggedright\arraybackslash}p{12cm} >{\raggedright\arraybackslash}p{1.5cm}}
        \toprule
        \textbf{System} & \textbf{SLOC*} & \textbf{Failure Types (\#)} & \textbf{\# Issues}\\
        \midrule
        \multirow{2}{*}{MapReduce} & \multirow{2}{*}{721K} & Hangs (7), Inconsistent state (5), Incorrect result (5), Resource exhaustion (2), Resource leak (1), Unexpected termination (6)& \multirow{2}{*}{26} \\ 
        \multirow{2}{*}{HDFS} & \multirow{2}{*}{706K}& Data loss (1), Hangs (1), Inconsistent state (1), Incorrect result (8), Resource leak (1), Unexpected termination (3) &  \multirow{2}{*}{15} \\
        HBase & 912K & Data loss (4), Hangs (8), Inconsistent state (2), Incorrect result (2), Unexpected termination (8) &  24 \\
        Cassandra & 1.1M & Hangs (6), Inconsistent state (5), Incorrect result (20), Unexpected termination (2)& 33 \\
        ZooKeeper & 184K & Hangs (4), Incorrect result (2), Resource leak (2)& 8 \\ \hline
         \multirow{2}{*}{Total} & \multirow{2}{*}{-}  & Data loss (5), Hangs (26), Inconsistent state (13), Incorrect result (37), Resource exhaustion (2), Resource leak (4), Unexpected termination (19) & \multirow{2}{*}{106} \\
        \bottomrule
    \end{tabular}
    \begin{flushleft}
\footnotesize{*SLOC is calculated based on the latest version. The actual SLOC may vary depending on the version in which the issue occurred.}
\end{flushleft}
    \vspace{-15pt}
\end{table*}

\subsubsection{Collection} Our experiments utilize datasets from a previous study~\cite{ist}, collected from the JIRA~\cite{JIRA}, a public issue-tracking platform. As shown in Table~\ref{tab:subjects}, the systems from which these issues were collected range from 144K to 3M lines of code, with 11 to 15 years of development. These systems are distributed and include computing, storage, and configuration frameworks. All the issues we studied are distributed issues, meaning the failure propagation involves multiple nodes or components, which makes RCA particularly challenging. To make our analysis more rigorous, we excluded issues that lacked run-time logs and filtered out those that lacked a detailed root cause analysis upon the closure.

\subsubsection{Labelling}\label{sec:label} All the issues in our study have been fixed and thoroughly investigated~\cite{ist, leesatapornwongsa2016taxdc, gunawi2014bugs}. The discussions and patches related to these issues are publicly available, which facilitates the labeling process. Two experienced developers, both are contributors of at least one above system, undertook the labeling procedure jointly. 


To label \textit{root cause summary}, annotators \add{read and summarize} the root cause identified from the discussions in the issue report and previous studies~\cite{ist}. They then distill the diagnostic procedure from the discussion, refining the summary based on mitigation strategies and fixing items. \add{The goal of the summary is to provide a clear and concise guide for maintainers to diagnose the issue.} For \textit{root cause localization}, due to differences in the granularity~\cite{yu2023nezha,lin2016log,li2021practical}, we standardized the process by identifying all components mentioned within the discussion and postmortem studies, allowing all approaches to select from them. \add{These components include code-related elements such as specific classes (e.g., JobClient), as well as non-code-related components including system resources (e.g., network, RPC) and abstract concepts (e.g., race conditions, deadlocks). The ground truth for root cause localization is derived from the studied root cause and the corresponding fixing patches, and it consists of one or two components.}



\subsection{Evaluation Metrics}

\subsubsection{Root Cause Summarization}

We use \textbf{BLEU}~\cite{papineni2002bleu}, \textbf{METEOR}~\cite{banerjee2005meteor}, and \textbf{ROUGE}~\cite{lin2004rouge} metrics, consistent with previous research~\cite{roy2024exploring, RCACopilot, wang2023rcagent}, to evaluate the quality of the generated root cause summarization from the aspect of text similarity with ground truth. We also leverage widely-used embedding models, OpenAI embedding~\cite{openai_embedding}, to embed the root cause summary and
 calculate the \textbf{Semantics} similarity.

Following previous works~\cite{wang2021automatic, saha2022mining, kang2024quantitative}, we also employ human evaluation
\noindent\textbf{Usefulness} indicates whether the summary accurately explains why the issue occurred and how to fix it. The same experts who annotated the datasets conducted this evaluation, ensuring consistency and a high level of expertise. They are offered the summary from all baselines alongside the corresponding ground truth for each issue. The evaluators are instructed to compare these results against the ground truth and rate the usefulness \add{for diagnosing the given issue} of each summary on a scale from 0 to 1. We then calculate the average score among evaluators for certain baseline.


\subsubsection{Root Cause Localization} We use the following two metrics to evaluate the effectiveness of \name and baselines. 
\begin{itemize}
\item \textbf{\textit{Exact Match}} \add{measures whether the generated root cause components exactly match the ground truth set of components. Any mismatch, either fewer or more components than the ground truth, results in a score of 0.}
\item \textbf{\textit{Top-k}} \add{measures whether the ground truth components are completely included within the top-k (k=3, 5) most-likely root cause components generated by models. For this metric, we prompt the model to generate the k most likely root cause components.}
\end{itemize}



\subsection{Comparative Methods}~\label{sec: baseline}
We deliberately refrained from comparing \name with traditional~\cite{lin2016log,yuan2010sherlog} and AIOps methodologies~\cite{yu2023nezha,lee2023eadro,lin2020fast} that utilize fault-free data during the procedure. Typically, when issues are reported and tracked using platforms such as JIRA\cite{JIRA}, obtaining fault-free comparative data is not feasible~\cite{ist, RCACopilot}. Additionally, we excluded AutoFL~\cite{kang2024quantitative} because solely based on ticket, it was hard to reproduce the bugs~\cite{yuan2014simple} of distributed system and obtain the runtime coverage. 

We selected RCACopilot~\cite{RCACopilot} and ReAct~\cite{roy2024exploring} as our primary baselines because of their similar RCA procedures and capabilities in handling both localization and summarization tasks. Both models will retrieve the historical issue reports as knowledge sources to enhance the performance of root cause analysis. Since neither model is open-source, we \textbf{reimplemented} their methods based on the corresponding papers. We also included the base model with five fixed examples to demonstrate the task for comparison. All methods utilize the \textbf{same backbone LLM} to ensure a fair comparison.


\subsection{Implementation Details}
The static analysis module in \name has been implemented using approximately 4,652 lines of code, written in both Java and Python. This development employs Soot~\cite{vallee2010soot} and Eclipse JDT Core~\cite{jdt_core}, facilitating joint analysis of both Java bytecode and source code. The experiments of \name and all baselines were conducted on a Linux machine (Ubuntu LTS 20.04).

For GPT-4o, \add{GPT-3.5~\cite{chatgpt}}, LLaMa-3.1~\cite{llama31}, \add{Gemini-1.5-Pro~\cite{gemini}} and Claude-3.5~\cite{anthropic}, we use the public APIs provided by OpenAI~\cite{chatgpt}, DeepInfra~\cite{Deepinfra}, \add{Google} and Anthropic~\cite{anthropic}, corresponding to the model \textit{GPT-4o-0513}, \add{\textit{GPT-3.5turbo-0125}}, \textit{LLaMa-3.1-405b}, \add{\textit{Gemini-1.5-Pro~\cite{gemini}}} and \textit{Claude-3.5-Sonne}, respectively. We set the \textit{temperature} of all models as zero to ensure the reproduction. The default number of sampled examples is set to 5 for all approaches. \add{The depth of analyzed constructed execution path is 2.}

%% file: sections/05_results.tex
\section{EVALUATION RESULTS}\label{sec: evaluation}

\begin{table*}[t]
    \centering
    \footnotesize
    \caption{Root cause analysis results from both ~\textit{summarization} and ~\textit{localization} dimensions for each system.}
    \label{tab:rq1-result}
        \begin{tabular}{l|l||ccccc|ccc}
            \toprule
               \multirow{2}{*}{\textbf{System}} & \multirow{2}{*}{\textbf{Model}} &
                \multicolumn{5}{c}{\textbf{Root Cause Summarization}} &
                \multicolumn{3}{c}{\textbf{Root Cause Localization}} \\
            \cmidrule{3-10}
                & &
                \textbf{BLEU-4}    &
                \textbf{ROUGE-1} &
                 \textbf{METEOR}  &
                \textbf{Semantics} &
                \textbf{Usefulness} &
                \textbf{Exact Match} &
                \textbf{Top-3} &
                \textbf{Top-5} \\
            \midrule
               \multirow{5}{*}{MapReduce} &  Base model& 0.147 & 0.451 & 0.354 & 0.781 & 0.645 & 0.346 &  0.615 & 0.692 \\
                & RCACopilot & 0.174 & 0.503 & 0.396 &  0.812 & 0.740 &  0.346  & 0.692 & 0.808 \\
                & ReAct & 0.172 & 0.494 &  0.384 & 0.804 & 0.683 &   0.385 & 0.654 & 0.808 \\

               & \name & \textbf{0.203} & \textbf{0.525} & \textbf{0.445} & \textbf{0.867} & \textbf{0.795} & \textbf{0.423} & \textbf{0.731} & \textbf{0.923}  \\
                        \midrule
               \multirow{5}{*}{HDFS} &  Base model & 0.145 &0.430  & 0.346 &0.787  & 0.627  & 0.200& 0.400 & 0.600 \\
                & RCACopilot & 0.196 & 0.479 & 0.407 & 0.835  & 0.650  & 0.400   & 0.667 & 0.800 \\
                & ReAct & 0.192 &  0.486 & 0.413 & 0.843 & 0.733&  0.400  & 0.533& 0.733\\

               & \name & \textbf{0.219} & \textbf{0.501} & \textbf{0.454} & \textbf{0.883} & \textbf{0.754} & \textbf{0.467} & \textbf{0.800} & \textbf{0.800}  \\
                        \midrule
               \multirow{5}{*}{HBase} &  Base model & 0.167& 0.487  & 0.385  &  0.817 & 0.683 & 0.250 & 0.542 & 0.917 \\
                & RCACopilot & 0.171 & 0.505 & 0.408& 0.873 & 0.734& 0.292 & 0.458 & 0.958\\
                & ReAct & 0.185 & 0.496 & 0.416& 0.844 & 0.729 &  0.375  & 0.458 & 0.917 \\

               & \name & \textbf{0.200} & \textbf{0.526} & \textbf{0.435}  & \textbf{0.893} & \textbf{0.745} & \textbf{0.417} & \textbf{0.667} & \textbf{0.958}  \\
                        \midrule
               \multirow{5}{*}{Cassandra} &  Base model & 0.149 & 0.451 & 0.340  & 0.784 & 0.648 & 0.364 & 0.576 & 0.697 \\
                & RCACopilot & 0.153 & 0.466 & 0.392 & 0.807 & 0.661 &   0.424 & 0.697 & 0.879 \\
                & ReAct & 0.165 & 0.484 & 0.407& 0.796 & 0.691 &  0.515 & 0.727 & 0.848 \\

               & \name & \textbf{0.208} & \textbf{0.522}  & \textbf{0.441} & \textbf{0.877} & \textbf{0.830} & \textbf{0.606} & \textbf{0.848} & \textbf{0.909}  \\
                        \midrule
               \multirow{5}{*}{ZooKeeper} &  Base model & 0.129 &0.429  & 0.333 & 0.756 & 0.588 & 0.250 & 0.500 & 0.875 \\
                & RCACopilot & 0.144 & 0.468 & 0.357 & 0.863 & 0.613&   \textbf{0.375} & 0.750 & 1.000\\
                & ReAct & 0.136& 0.438 & 0.342 & 0.828 & 0.594 &   0.250 & 0.625 & 0.875\\
               & \name & \textbf{0.185} & \textbf{0.497}  & \textbf{0.381} & \textbf{0.894} & \textbf{0.625} & 0.250 & \textbf{0.875} & \textbf{1.000}  \\
            \bottomrule
            \end{tabular}%

\end{table*}

\subsection{RQ1: Effectiveness of \name}

To evaluate the performance of \name in RCA tasks, we conduct a thorough evaluation with comparison to other baselines. The results of this evaluation are detailed in Table~\ref{tab:rq1-result} and all approaches employ \textit{GPT-4o-0513} as the uniform backbone model. We analyze the evaluation results from two dimensions: \textit{Root Cause Summarization} and \textit{Root Cause Localization}. 

\subsubsection{Root Cause Summarization} We compare \name with baselines in terms of root cause summarization from the aspects of syntax similarity, semantic similarity, and human-evaluated metrics. Regarding syntax similarity, we observe that \name outperforms the best performing approach, RCACopilot, showing improvements of 22.0\% in \textit{BLEU-4}, 6.8\% in \textit{ROUGE-1}, and 10.6\% in \textit{METEOR} scores, on average. These enhancements are even more pronounced in specific system: for instance, when analyzing issues from ZooKeeper, \name surpasses the best-performing approach by 28.5\% in \textit{BLEU-4}. For semantic similarity,  \name consistently outperforms the best-performing baseline with improvements ranging from 2.3\% to 8.7\% across all evaluated systems, providing another dimension of effectiveness against the baselines. In terms of \textit{Usefulness}, which could directly measure the help for the maintainers, \name exhibits substantial enhancements. This underscores the practical impact of integrating code knowledge in real-world distributed system maintenance.

\subsubsection{Root Cause Localization}  According to the evaluation results, it is clear that \name outperforms all baselines in faulty components localization. \name outperforms the best performing baseline in all evaluated systems across almost all metrics. Specifically, when averaged across all systems, \name outperforms RCACopilot by 28.3\%, 20.7\% and 4.3\% for metric \textit{Exact Match}, \textit{Top-3} and \textit{Top-5}, respectively. Notably, despite utilizing the same foundational model (\ie GPT-4o), \name achieves a remarkable 56.3\% improvement of the base model in \textit{Exact Match}, the most challenging metric, underscoring the effectiveness in the task of root cause localization. These results underscore the importance of integrating internal system knowledge (\ie code) over relying solely on historical data (\eg ReAct) in RCA tasks.

\begin{tcolorbox}[boxsep=1pt,left=2pt,right=2pt,top=3pt,bottom=2pt,width=\columnwidth,colback=white!95!black,boxrule=1pt, colbacktitle=white!30!black,toptitle=2pt,bottomtitle=1pt,opacitybacktitle=0.4]
\textbf{Answer to RQ1.} By incorporating the code knowledge into the procedure of RCA, \name demonstrates superior performance in both dimensions of root cause summarization and localization, significantly outperforms all baselines and especially the base model.
\end{tcolorbox}

\subsection{RQ2: Impact of different phases in \name} 

\begin{table*}[tbp]
    \centering
    \scriptsize
    \caption{Ablation Study of \name.}
    \label{tab:rq2}
        \begin{tabular}{l|l||ccccc|ccc}
            \toprule
               \multirow{2}{*}{\textbf{System}} & \multirow{2}{*}{\textbf{Setting}} &
                \multicolumn{5}{c}{\textbf{Root Cause Summarization}} &
                \multicolumn{3}{c}{\textbf{Root Cause Localization}} \\
            \cmidrule{3-10}
                & &
                \textbf{BLEU-4}    &
                \textbf{ROUGE-1} &
                 \textbf{METEOR}  &
                \textbf{Semantics} &
                \textbf{Usefulness} &
                \textbf{Exact Match} &
                \textbf{Top-3} &
                \textbf{Top-5} \\
            \midrule
               \multirow{5}{*}{All} & \name  & \textbf{0.205} & \textbf{0.519} & \textbf{0.438} &  \textbf{0.880} &\textbf{0.776} & \textbf{0.472} &  \textbf{0.774} & \textbf{0.915} \\
                & w/o Logging Source Retrieval & 0.145 & 0.462 & 0.338 & 0.815 &  0.660 & 0.387 & 0.557 & 0.660 \\
                & w/o Execution Path Reconstruction & 0.171 & 0.472  &  0.403 & 0.827 & 0.685 &  0.349  & 0.632 & 0.821 \\
                & \add{w/ 1-Degree Execution Path Reconstruction} & \add{0.189} & \add{0.507} & \add{0.373} & \add{0.872} & \add{0.723} & \add{0.443} & \add{0.698} & \add{0.842} \\
& \add{w/ 3-Degree Execution Path Reconstruction} & \add{0.201} & \add{0.524} & \add{0.419} & \add{0.891} & \add{0.693} & \add{0.349} & \add{0.745} & \add{0.906} \\
                & w/o Failure-Related Code Profiling & 0.175 & 0.507  & 0.399 & 0.847  & 0.734  &  0.396  & 0.679 & 0.792 \\
                \midrule
                & \add{only w/ Full JIRA Discussion} & \add{0.495} & \add{0.712} & \add{0.775} & \add{0.914} & \add{0.895} & \add{0.849} & \add{0.906} & \add{0.934} \\
            \bottomrule
            \end{tabular}%
            \vspace{-10pt}

\end{table*}

To evaluate the individual contribution and parameter setting of each phase within the \name, we conducted an ablation study by creating three variants of \name, each missing a different phase. Specifically, to evaluate the framework without the logging source retrieval phase, we adopted a new code-enhanced method that retrieves the top 5 methods (calculated by OpenAI embedding~\cite{openai_embedding} similarity) from the project codebase for comparison. \add{Additionally, to estimate the impact of analyzed execution depth, we also investigated execution depth parameters ranging from 1 to 3.}

Table~\ref{tab:rq2} demonstrates the experimental results. The results indicate that without conducting the logging source retrieval phase, the overall performance of \name generally declines across all metrics. There is a decrease of 18\% in \textit{Exact Match}, while \textit{METEOR} and \textit{Usefulness} fall by 22.9\% and 14.9\% respectively. More critically, some metrics (\eg \textit{METEOR}, \textit{Top-5}) even fall below those of the raw LLM. This suggests that when the retrieved code snippets are irrelevant to the failure at hand, they can potentially mislead the model. Additionally, when execution path reconstruction is not conducted, the overall performance of \name also deteriorates significantly. The \textit{BLEU-4} score drops from 0.205 to 0.171 and the \textit{Usefulness} score from 0.776 to 0.685, representing decreases ranging from 16.6\% to 11.7\%, respectively. It is noteworthy that the \textit{Exact Match} performance is even lower than search-based method. \add{This indicates that without analyzing the execution paths and relying solely on the log surrounding code snippets, it is challenging for the LLM to accurately identify the root cause. As shown by our experiments with different execution depths, where each invocation is treated as one degree, 1-degree paths may miss critical code
information, while 3-degree paths may introduce too much irrelevant code. The experiment results suggest that 2-degree paths (default setting of \name) provide optimal performance in this scenario.} \add{Additionally, regarding RPC invocations, approximately 15.7\% of the invocations in a reconstructed execution path are RPC invocations, demonstrating the effectiveness of proposed RPCBridge.}

Furthermore, when the phase of failure-related code profiling is omitted, the performance of \name for root cause localization drops obviously (16.1\%, reflected by \textit{Exact Match}), while the performance of root cause summarization remains relatively high. Furthermore, it is puzzling to observe that the \textit{Top-5} accuracy remains at a similar level with \textit{Top-3} accuracy. Upon investigating this phenomenon, we discovered that it is due to extremely long code contexts. These results underscore the critical role of the failure-related code profiling phase.

\begin{tcolorbox}[boxsep=1pt,left=2pt,right=2pt,top=3pt,bottom=2pt,width=\columnwidth,colback=white!95!black,boxrule=1pt, colbacktitle=white!30!black,toptitle=2pt,bottomtitle=1pt,opacitybacktitle=0.4]
\textbf{Answer to RQ2.} The ablation study indicates that the elimination of any component notably reduces the overall performance. Particularly, the results demonstrate that merely searching the codebase without the \name framework falls short of expectations, emphasizing that each component contributes to the overall performance.
\end{tcolorbox}

\subsection{RQ3:  Generalizability of \name}

\begin{table*}[tbp]
    \centering
    \footnotesize
    \caption{The performance of \name with different backbone models.}
    \label{tab:rq3}
        \begin{tabular}{l|l||ccccc|ccc}
            \toprule
               \multirow{2}{*}{\textbf{System}} & \multirow{2}{*}{\textbf{Framework}} &
                \multicolumn{5}{c}{\textbf{Root Cause Summarization}} &
                \multicolumn{3}{c}{\textbf{Root Cause Localization}} \\
            \cmidrule{3-10}
                & &
                \textbf{BLEU-4}    &
                \textbf{ROUGE-1} &
                 \textbf{METEOR}  &
                \textbf{Semantics} &
                \textbf{Usefulness} &
                \textbf{Exact Match} &
                \textbf{Top-3} &
                \textbf{Top-5} \\
            \midrule
               \multirow{3}{*}{GPT-4o} &  Base & 0.151 & 0.455 & 0.354 & 0.789 &  0.648 & 0.302 & 0.547  & 0.745  \\
                & \name & 0.205 & 0.519 & 0.438 & 0.880 &  0.776 & 0.472 & 0.774 & 0.915\\
                & $\Delta$ & $\uparrow35.8\%$ & $\uparrow14.1\%$ & $\uparrow23.7\%$ & $\uparrow11.5\%$ & $\uparrow19.8\%$  & $\uparrow56.3\%$ & $\uparrow41.4\%$ & $\uparrow22.8\%$ \\
                \midrule
\multirow{3}{*}{\add{GPT-3.5}} & \add{Base} & \add{0.124} & \add{0.408} & \add{0.309} & \add{0.755} & \add{0.541} & \add{0.264} & \add{0.472} & \add{0.632} \\
& \add{\name} & \add{0.157} & \add{0.461} & \add{0.361} & \add{0.802} & \add{0.592} & \add{0.368} & \add{0.557} & \add{0.764} \\
& \add{$\Delta$} & $\add{\uparrow26.6\%}$ & $\add{\uparrow13.0\%}$ & $\add{\uparrow16.8\%}$ & $\add{\uparrow6.22\%}$ & $\add{\uparrow9.4\%}$ & $\add{\uparrow39.4\%}$ & $\add{\uparrow18.0\%}$ & $\add{\uparrow20.9\%}$ \\
                        \midrule
                \multirow{3}{*}{LLaMa-3.1-405b} &  Base & 0.150  & 0.456  & 0.348 &  0.764 &  0.626 & 0.283 &  0.557 & 0.811 \\
                & \name & 0.189 & 0.514 & 0.427 & 0.838 &  0.702  & 0.434 & 0.745 & 0.877\\
                & $\Delta$ &$\uparrow26.0\%$ &$\uparrow12.7\%$  &$\uparrow22.7\%$ &  $\uparrow9.7\%$&  $\uparrow12.1\%$ & $\uparrow53.3\%$ & $\uparrow33.9\%$&$\uparrow8.1\%$ \\
                \midrule
                \multirow{3}{*}{Claude-3.5-Sonnet} &  Base & 0.107 & 0.441 & 0.337&  0.800 & 0.652 & 0.349 &  0.660 & 0.868 \\
                & \name & 0.148 & 0.470 & 0.419& 0.854 &  0.744 & 0.453 & 0.764 & 0.943 \\
                & $\Delta$ &$\uparrow38.3\%$ &  $\uparrow6.6\%$& $\uparrow24.3\%$& $\uparrow6.8\%$ & $\uparrow14.1\%$   & $\uparrow29.7\%$ & $\uparrow15.7\%$ & $\uparrow8.7\%$ \\
                        \midrule
 \multirow{3}{*}{\add{Gemini-1.5-Pro}} & \add{Base} & \add{0.161} & \add{0.460} & \add{0.361} & \add{0.794} & \add{0.633} & \add{0.321} & \add{0.528} & \add{0.764}\\
& \add{\name} & \add{0.224} & \add{0.523} & \add{0.408} & \add{0.859} & \add{0.737} & \add{0.443} & \add{0.736} & \add{0.887}\\
& \add{$\Delta$} & $\add{\uparrow39.1\%}$ & $\add{\uparrow13.7\%}$ & $\add{\uparrow13.3\%}$ & $\add{\uparrow8.2\%}$ & $\add{\uparrow16.4\%}$ & $\add{\uparrow38.0\%}$ & $\add{\uparrow39.4\%}$ & $\add{\uparrow16.1\%}$ \\
            \bottomrule
            \end{tabular}%
                        \vspace{-10pt}
\end{table*}

In this RQ, we evaluate the performance of \name by utilizing various LLMs in conjunction with our framework. We have selected three representative LLMs that are frequently used in research, specifically \textit{GPT-3.5}, \textit{GPT-4o}, \textit{Llama-3.1-405b}, \textit{Gemini-1.5-Pro}, and \textit{Claude-3.5-Sonnet}. We selected versions of these models with enough parameter size to ensure the capability of complex prompt understanding~\cite{xu2023prompting, gao2023what}. 

The experimental results are shown in Table.~\ref{tab:rq3}. 
Our observations indicate that the \name framework can consistently improve the performance of all utilized base models in terms of all metrics by a large margin. On average, all models have been improved \add{by 43.3\%, 29.7\% and 15.3\%}
in localizing the root causes based on the metrics of \textit{Exact Match}, \textit{Top-3}, and \textit{Top-5} respectively. Furthermore, when it comes to summarizing the root causes, we observed an average improvement of \add{33.2\% (as reflected by \textit{BLEU-4}), 20.2\% (as reflected by \textit{METEOR}), and 14.4\% (as reflected by \textit{Usefulness})}. Particularly noteworthy is the performance of the \textit{Claude-3.5-Sonnet} model within the \name framework, which surpasses the raw model in \textit{Top-3}. The experiments compared the performance of the latest and most powerful LLMs in root cause analysis tasks. The results indicate that although \textit{GPT-4o} is not the most advanced raw model from some aspects (\eg \textit{Exact Match}), the enhancements provided by \name are still the most significant.



The results not only demonstrate the advantage of \name's design but also demonstrate the generalizability for different backbone models. We believe that the performance of \name can be further improved with the development of LLMs.

\add{Furthermore, the experiments with full JIRA discussion in Table.~\ref{tab:rq2} further demonstrate \name's generalizability. The results show dramatic improvements after incorporating detailed JIRA issue discussions, indicating that those LLMs have limited memorization of JIRA issue results. Moreover, Table.~\ref{tab:rq3} shows that all five LLMs perform moderately when provided with only the raw issues, which more closely resembles their training data format~\cite{huang2023not}. However, after incorporating the knowledge provided by COCA, all models exhibit consistent and significant improvements in performance.}

\begin{tcolorbox}[boxsep=1pt,left=2pt,right=2pt,top=3pt,bottom=2pt,width=\linewidth,colback=white!95!black,boxrule=1pt, colbacktitle=white!30!black,toptitle=2pt,bottomtitle=1pt,opacitybacktitle=0.4]
\textbf{Answer to RQ3.} \name consistently enhances the performance of both root cause summarization and localization, regardless of the type of backbone models employed. \add{Moreover, our experiments demonstrate the limited data leakage, further highlighting the generalizability of \name}.
\end{tcolorbox}

%% file: sections/06_discussion.tex
\section{Discussion}

\subsection{Case Study}

\begin{figure}[t]
    \centering
    \includegraphics[width=0.95\columnwidth]{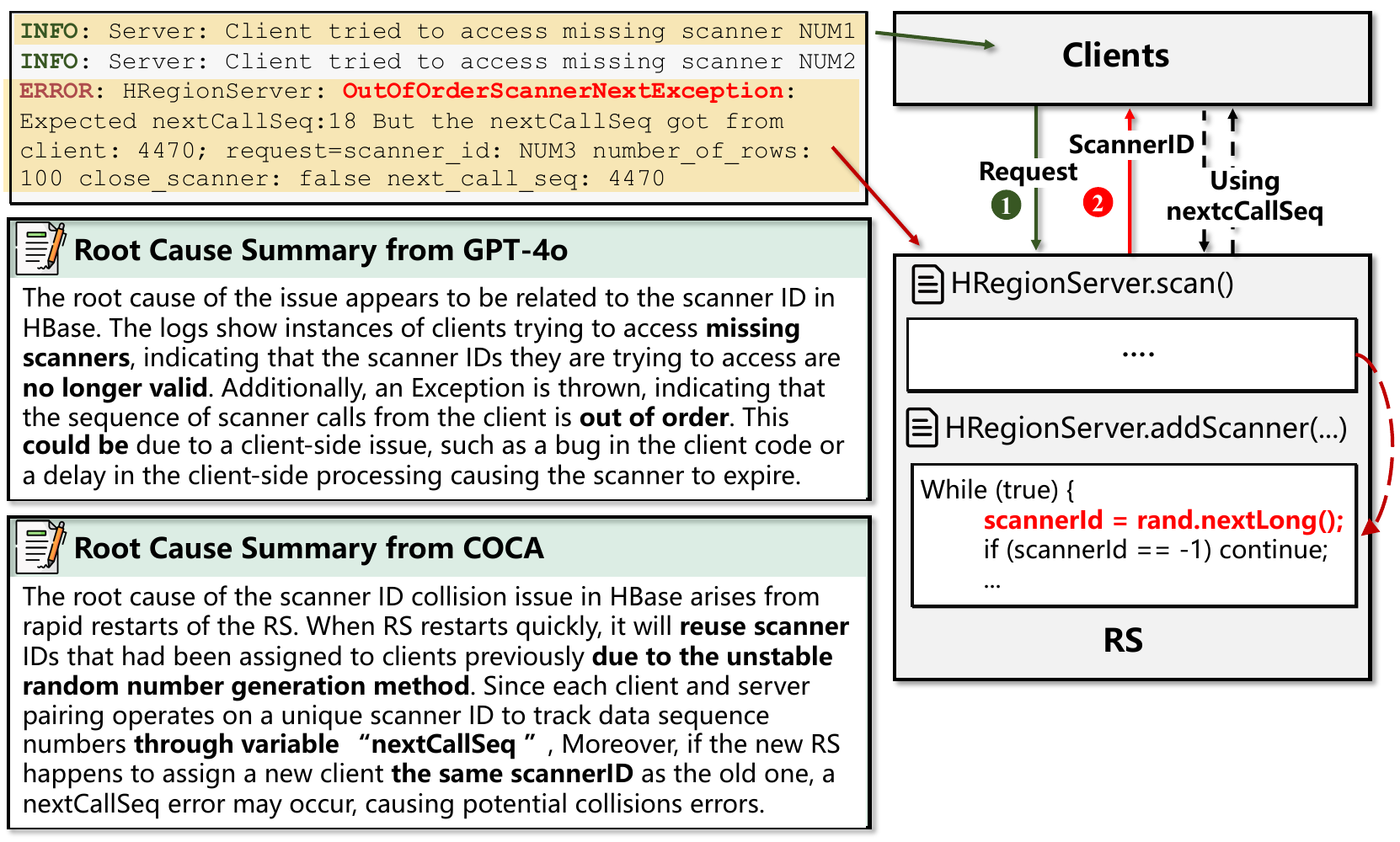}
    \vspace{-0.05in}
    \caption[Case study of HBASE-9821]{Case study of HBASE-9821\protect\footnotemark.}
    \label{fig: case-study}
    \vspace{-0.05in}
\end{figure}
\footnotetext{\label{myfootnote}\url{https://issues.apache.org/jira/browse/HBASE-9821}}

To demonstrate the practical value of \name for system maintainers, we analyze HBASE-982\textsuperscript{\ref{myfootnote}} as a practical example. This case illustrates how \name significantly improves the efficiency and accuracy of root cause analysis for system maintainers. The issue involves scanner ID collisions in the HBASE system, where multiple clients attempt to access non-existent scanner IDs, resulting in \textit{nextCallSeq} mismatch errors. 

\add{While GPT-4o provides a surface-level diagnosis based solely on the issue report and logs, identifying only that the scanner IDs are invalid and clients are unaware of this condition, \name delivers substantially more actionable insights for maintainers by leveraging code knowledge extracted from logs and traces.} \name determines the root cause of the issue is the reuse of scanner IDs, then concludes that scanner IDs are generated through random numbers by reviewing the code. When an HRegionServer (RS) quickly restarts, it is treated as a new RS and is assigned the same scanner ID as before due to the unreliable random number generation logic without dynamic seeds. Consequently, this leads to inconsistencies of scanner IDs for clients, resulting in a system crash.

Compared to solely relying on issue reports, the integration of code knowledge can substantially enhance the information available to the LLM prior to making an inference, thereby generating a more comprehensive and detailed RCA result. 

\add{To demonstrate \name's benefits for maintainers, we conducted a controlled experiment of this case with an experienced annotator, who is also the contributor to HBase. When presented with GPT-4o's RCA summarization, the annotator spent over 3 hours analyzing the system's execution path and related code based on the traces and logs available in the issue report, which \name had already automatically summarized. Notably, this issue took a full day of active discussion on the JIRA forum before reaching the conclusion. This real-world example demonstrates how \name accelerates the diagnosis process by automatically reconstructing and analyzing system execution paths with LLM that would otherwise require hours of manual investigation by skilled maintainers.}

\subsection{\add{Practicality of \name}}
\add{\name is designed to facilitate the diagnosis of issues and incidents~\cite{chen2020towards, RCACopilot} in real-world distributed systems, such as modern cloud systems (e.g., Azure and AWS). These mature systems have employed issue-tracking platforms (such as ICM~\cite{chen2020towards} in Microsoft, JIRA used by Apache) to streamline issue management and enhance system reliability.}

\add{When an incident or issue is reported, \name will outperform current methods in three aspects. In terms of \textbf{effectiveness}, our extensive experiments on real-world JIRA issues demonstrate \name's strong performance in production environments. The evaluation results show high accuracy in root cause analysis, consistent performance across different types of issues, which is the strong evidence of \name's practical capability in real-world systems. Additionally, \name's \textbf{extensibility} allows for seamless adaptation to various issue tracking systems without the need to reproduce the issue. These platforms share similar settings with user-reported issues, including descriptions, logs, and traces. Besides, for \textbf{efficiency}, it typically takes maintainers several hours to days (based on average response time from JIRA issue maintenance records). However, \name provides RCA results with an average response time of \textit{19.4} seconds while automatically identifying and analyzing an average of \textit{26.9} methods from execution paths in hundreds of lines of logs and stack traces.}


\subsection{Threats to Validity}
\noindent
\textbf{Potential Data Leakage.}
One of the main concerns of this work is the potential data leakage issue due to the utilization of public issue reports. Specifically, the target issue report may be trained during the pretraining or retrieved by our sampling algorithm, leading it to memorize rather than infer~\cite{yang2023code, li2023exploring, huang2023not} the result. However, this risk may be mitigated by several factors.
Firstly, many of the root causes are not available on JIRA, which reduces the chances of data leakage. Secondly, our complex prompt format is unlikely to appear directly in the training dataset, especially compared to raw issue reports, \add{as the experiment results shown in Table~\ref{tab:rq3} for 5 representative LLMs}. \add{Thirdly, as shown in Table~\ref{tab:rq2}, the performance improves significantly after feeding the raw JIRA issue discussions, demonstrating that those LLMs have limited memorization of RCA results.} Moreover, during the dataset collection phase, we manually reviewed each issue's discussion and linking relationships to ensure that there were no duplicate issue reports with the same root cause.

\noindent \textbf{The Precision of Executed Path Reconstruction.}
Static analysis has inherent limitations in its analysis boundary\cite{LiSecC22, Wang2024ISSTA, SamhiICSE22} and struggles with dynamic bindings\cite{Li2019TOSEM, Bruce2020JShrink, Grech2017PTaint}. This affects its accuracy in constructing call graphs, especially when dealing with network interactions and the use of reflection, proxies and multi-threading in distributed systems. To mitigate this, we propose and implement a novel RPC bridging method to integrate RPCs within \name to enhance the its analytical capabilities. In addition, ablation studies also show that compared with similarity-based code base retrieval methods, reconstructing execution paths can maximize the introduction of fault-related code fragments and restore the system execution state as much as possible.

%% file: sections/07_related_work.tex
\section{Related Work}\label{sec: related}

\textbf{Monitor Data-driven RCA.} Most of the existing RCA approaches~\cite{chen2014causeinfer, amar2019mining, lin2016log, rosenberg2020spectrum, lin2020fast, lee2023eadro,yu2023nezha, li2022intelligent,jiang2025l4} only use monitor data (e.g., logs or traces) as input. They typically compare whether the runtime data pattern changes under normal and abnormal system states to figure out the root cause. For instance, LogCluster\cite{lin2016log} utilizes a clustering algorithm to collect log clusters, which are then employed to discern log abnormal patterns by comparing them with test log sequences. NeZha~\cite{yu2023nezha} contrasts event graphs from fault-free and fault-affected systems to deduce the root causes. However, this type of approach has three limitations: (1) they rely on the normal runtime data pattern of the system, which is not readily available in practice. (2) they usually only output the root cause module and lack a root cause summarization. (3) they do not incorporate related code as contextual information, leading to underperforming diagnostic results.

\textbf{LLM-based RCA.} The recent advent of LLMs has facilitated the proposal of several LLM-based RCA methodologies~\cite{RCACopilot,roy2024exploring,kang2024quantitative,shan2024face}, since LLMs inherently possess domain knowledge of systems and corresponding failures. RCACopilot~\cite{RCACopilot}, which was introduced as an LLM-based RCA approach equipped with retrieval tools. By analyzing historical incidents, ReAct~\cite{roy2024exploring} can understand complex system issues, thereby continuously enhancing its performance. 

Compared to existing LLM-based RCA approaches, \name extends its diagnostic capabilities beyond runtime information and historical bug reports by incorporating code knowledge. \add{\name can retrieve and analyze internal system knowledge (\ie code) without requiring issue reproduction to obtain code coverage during execution, which fault localization methods did~\cite{kang2024quantitative} but is impractical in distributed systems~\cite{yuan2014simple,yuan2010sherlog}. From the perspective of fault types, \name address a variety of issues, including both code bugs and non-code problems (e.g., resource limitations, misconfigurations, or environmental factors) by providing code knowledge to enhance the comprehension of the system, thereby facilitating the RCA process, while fault localization methods focus solely on localizing reproducible code bugs. Regarding analysis scope, \name performs root cause analysis by summarizing and localizing runtime problems. In contrast, fault localization techniques are restricted to localizing faults within the code itself, without addressing non-code or environmental factors. }Thus, \name overcomes the limitations of previous LLM-based approaches by providing a more thorough and contextualized RCA method.

%% file: sections/08_conclusion.tex
\section{Conclusion}
In this paper, we propose \name, the first work incorporating code knowledge into an automatic root cause analysis framework for issue reports in distributed systems. Experimental results show that \name outperforms all baselines and can be generalized to various LLMs. \name demonstrates its promising future in the context of the evolving LLM techniques, elevating the upper limit of LLM's capability in RCA task and benefiting both researchers and maintainers. 


\section*{Acknowledgment}
The work described in this paper was supported by the Research Grants Council of the Hong Kong Speci al Administrative Region, China (No. CUHK 14206921 of the General Research Fund), and RGC Grant for Theme-based Research Scheme Project (RGC Ref. No. T43-513/23-N), and the National Nature Science Foundation of China under Grant (No. 62402536).